\documentclass[12pt]{article}

\usepackage{latexsym}

\font\tenmsbm=msbm10 scaled 1200
\font\sevenmsbm=msbm9
\newfam\msbmfam
\textfont\msbmfam=\tenmsbm
\scriptfont\msbmfam=\sevenmsbm

\makeatletter
\@addtoreset{equation}{section}
\makeatother

\def\beq{\begin{equation}}
\def\eeq{\end{equation}}
\def\bea{\begin{eqnarray}}
\def\eea{\end{eqnarray}}
\def\bet{\begin{tabular}}
\def\eet{\end{tabular}}

\catcode`@=11
\def\quad@rato#1#2{{\vcenter{\vbox{
        \hrule height#2pt
        \hbox{\vrule width#2pt height#1pt \kern#1pt \vrule width#2pt}
        \hrule height#2pt} }}}
\def\quadratello{\mathchoice
\quad@rato5{.5}\quad@rato5{.5}\quad@rato{3.5}{.35}\quad@rato{2.5}{.25} }

\begin{document}

\begin{titlepage}

\begin{flushright}
La Plata-Th 00/11\\ November 2000\\
\end{flushright}

\vspace{1truecm}

\begin{center}

{\Large \bf An alternative formulation of classical\\ \vskip 0.3
cm electromagnetic duality}

\vspace{2cm}

Kang Li$^{a,b, }$\footnote{kangli@mail.hz.zj.cn} and Carlos M.
Na\'on $ ^{b,}$\footnote{naon@venus.fisica.unlp.edu.ar}

\vspace{2cm}

{\it\small $^a$Department of Physics, Zhejiang University,
Hangzhou, 310027, P.R. China

\smallskip
$^b$Instituto de F\'{\i}sica La Plata, Departamento de F\'{\i}sica, Facultad de
Ciencias Exactas, Universidad Nacional de La Plata , CC 67,1900 La Plata, Argentina }

 \vspace{1cm}

\begin{abstract}
\vspace{0.5cm} By introducing a doublet of electromagnetic four dimensional vector
potentials, we set up a manifestly Lorentz covariant and $SO(2)$ duality invariant
classical field theory of electric and magnetic charges. In our formulation one does
not need to introduce the concept of Dirac string.

\end{abstract}

\end{center}
\vskip 0.5truecm
\noindent


\end{titlepage}
\newpage
\baselineskip 6 mm

\section{Introduction}

Recently there has been much interest in the study of electromagnetic (EM) duality,
perhaps due to its role in unifying theories \cite{Schwarz-Seiberg}. In particular,
the construction of a fundamental "M-theory" in eleven dimensions, based on extended
objects called branes, relies on the implementation of an EM duality. >From Maxwell's
equations we know that general EM duality implies the existence of magnetic charge
(monopole) and currents. However, when considering the quantum dynamics of particles
carrying both electric and magnetic charges (dyons) one faces the lack of a naturally
defined classical field theory despite of the fact that a consistent quantum field
theory does exist \cite{Brandt}. This issue was analyzed in recent contributions by
considering certain generalized Dirac-strings \cite{LM} (See also \cite{GM} for a
formulation of dual quantum electrodynamics based on the original string-dependent
action of Dirac and Schwinger). In \cite{Carneiro1} a non local Lagrangian formalism
was proposed whose quantized version allowed to compute probability amplitudes for
charge-monopole scattering \cite{Carneiro2}. There were also some other recent works
that focused on different aspects of classical electromagnetism such as its connection
to General Relativity \cite{Mendez}, consistency conditions \cite{SD} \cite{Galvao},
etc.

The aim of this work is to present and explore an alternative formulation of EM
duality, in a classical context, which avoids the concept of Dirac-string. This
classical field theory is based on the introduction of a doublet of 4-dimensional
vector potentials that have no singularity around the magnetic monopoles. It posses
manifest Lorentz covariance and $SO(2)$ duality symmetry. The main advantage of our
formulation is that, since the gauge fields are regular, one expects that their
quantization will be straightforward.

The paper is organized as follows. In the next section we give a
brief review of classical EM duality and explain why one needs to
introduce the Dirac monopole and the Dirac string around the
monopole. In the third section we describe in some detail our
formulation. In section $4$, in order to illustrate the physical
content of this model and allow comparison with the Dirac-string
formulation, we solve the classical equations of motion for some
specific static and non-static cases. In section $5$ we derive a
generalized Lorentz force formula and show how to get the
electric charge quantization condition. Finally, in section $6$
we display the Lagrangian form of our formulation and present our
main conclusions.

\section{EM duality and the Dirac string}

 Let us start by recalling the basic concepts related to EM duality and the
 Dirac monopole. In terms of the electric field $\mathbf{E}$, the magnetic induction
 $\mathbf{B}$, the charge and current densities $\rho_e$ and
 $\mathbf{j_e}$, Maxwell's equations read
 \begin{equation}
 \nabla\cdot\mathbf{E}=\rho_e,~~~~~~\nabla\times\mathbf{B}=\frac{\partial{\mathbf{E}}}{\partial{t}}
 +\mathbf{j_e}~,
\end{equation}
\begin{equation}
\nabla\cdot\mathbf{B}=0,~~~~~~~~~~\nabla\times\mathbf{E}=-\frac{\partial{\mathbf{B}}}{\partial{t}}.
\end{equation}
where we set $c=\hbar=1$ and $ \mu_0=\varepsilon_0=1$. As usual,
one introduces the electric scalar potential $\phi_e$ and the
magnetic vector potential $\mathbf{A_m}$ such that
\begin{equation}
\mathbf{E}=-\nabla\phi_e -\frac{\partial \mathbf{A_m}}{\partial t}
\end{equation}
\begin{equation}
\mathbf{B}=\nabla\times\mathbf{A_m}
\end{equation}
\begin{equation}
A_\mu=(\phi_e,-\mathbf{A_m}),~~~x_\mu=(t,-\mathbf{x}),~~~J^e_\mu=(\rho_e,-\mathbf{j_e}).
\end{equation}
Thus, in terms of the electro-magnetic field tensor
\begin{equation}
F_{\mu\nu}=\partial_\mu A_\nu-\partial_\nu A_\mu ,
\end{equation}
Maxwell's equations (2.1) and (2.2) can be written in a manifestly Lorentz covariant
form as
\begin{equation}
\partial^\mu F_{\mu\nu}=J^e_\nu ,
\end{equation}
and
\begin{equation}
\partial_\sigma F_{\mu\nu}+\partial_\mu F_{\nu\sigma}+\partial_\nu
F_{\sigma\mu}=0.
\end{equation}
Furthermore, if we define the dual of the electromagnetic field tensor as
\begin{equation}
{^\ast} F_{\mu\nu}=\frac{1}{2}\epsilon_{\mu\nu\rho\sigma}F^{\rho\sigma},
\end{equation}
Maxwell's equations become
\begin{equation}
\partial^\mu F_{\mu\nu}=J^e_\nu,~~~~~~ \partial^\mu
\,{^\ast} F_{\mu\nu}=0.
\end{equation}

In the source free case, i.e. $J^e_\mu=(\rho_e,-\mathbf{j_e})=0$,
the duality symmetry of the above equations becomes apparent.
Indeed, it is easy to show that under the dual transformation
given by $F_{\mu\nu}\rightarrow{^\ast} F_{\mu\nu} ,~~{^\ast}
F_{\mu\nu}\rightarrow{^\ast}{^\ast} F_{\mu\nu}=-F_{\mu\nu}$, the
equations (2.10) are unchanged. In fact, taking into account that
\begin{equation}
F_{\mu\nu}=\left( \begin{array}{cccc} 0&E_x&E_y&E_z\\
-E_x&0&B_z&-B_y\\-E_y&-B_z&0&B_x\\-E_z&B_y&-B_x&0\end{array} \right),
\end{equation}
and
\begin{equation}
{^\ast} F_{\mu\nu}=\left( \begin{array}{cccc} 0&B_x&B_y&B_z\\
-B_x&0&-E_z&E_y\\-B_y&E_z&0&-E_x\\-B_z&-E_y&E_x &0\end{array} \right),
\end{equation}
one can readily convince oneself that the dual transformation is equivalent to
performing the following replacement in Maxwell's equations:
\begin{equation}
\mathbf{E}\rightarrow\mathbf{B},~~~~\mathbf{B}\rightarrow -\mathbf{E}.
\end{equation}

The duality symmetry  is immediately broken if a non-zero
electric current $J^e_\mu$ enters the theory, unless a non-zero
magnetic current $J^m_\mu =(\rho_m,-\mathbf{j_m})$ is also
introduced. When both electric and magnetic sources are included
Maxwell's equations read
\begin{equation}
 \nabla\cdot\mathbf{E}=\rho_e,~~~~~~\nabla\times\mathbf{B}=\mathbf{j_e}+\frac{\partial{\mathbf{E}}}{\partial{t}},
\end{equation}
\begin{equation}
\nabla\cdot\mathbf{B}=\rho_m,~~~~~~~~~~\nabla\times\mathbf{E}=-\mathbf{j_m}-\frac{\partial{\mathbf{B}}}{\partial{t}}
\end{equation}
and both electric and magnetic currents are conserved, i.e. they satisfy
\begin{equation}
\partial^\mu J^e_\mu=0
\end{equation}
\begin{equation}
\partial^\mu J^m_\mu=0.
\end{equation}

Maxwell's equations and currents conservation conditions above are obviously invariant
under the dual transformation, i.e. they are invariant under the replacement (2.13)
together with
\begin{equation}
\rho_e\rightarrow\rho_m,~~\rho_m\rightarrow
-\rho_e,~~\mathbf{j_e}\rightarrow\mathbf{j_m},~~\mathbf{j_m}\rightarrow -
\mathbf{j_e}.
\end{equation}

 At this point the question arises about how to formulate Maxwell's duality theory in the presence of
sources, in a Lorentz covariant way. It is clear that one needs to modify the
Maxwell's equation (2.10) in the form
\begin{equation}
\partial^\mu F_{\mu\nu}=J^e_\nu,~~~~~~ \partial^\mu
\,{^\ast} F_{\mu\nu}=J^m_\nu.
\end{equation}
But, as we know, if the field tensor $F_{\mu\nu}$ is defined by equation (2.6) then
the second equation of (2.10) (Bianchi identity) always holds unless the potential
$A_\mu$ have a singularity somewhere. In other words, if we compute the magnetic flux
through a sphere $S^2$ which surrounds the monopole with magnetic charge $g$, we shall
always get
\begin{equation}
\int\int_{s^2}\mathbf{B}\cdot d\mathbf{S}=\int\int_{s^2} (\nabla\times\mathbf{A})\cdot
d\mathbf{S}=0,
\end{equation}
although, due to the existence of the monopole, we must have a flux equal to $g$. So,
as is well-known, the way out of this problem is to allow $\mathbf{A}$ to have a
singularity somewhere on the sphere. This argument can be used for any radius of the
spheres surrounding the monopole, so by increasing it from zero to infinity we
conclude that the monopole has attached a line of singularities. This line is called
the Dirac string \cite{Dirac}, and this kind of monopole is also called Dirac
monopole. The Dirac string is not a physical observable and one should not be able to
measure it, so the orientation of the Dirac string  can be chosen arbitrarily
(different orientations correspond to different gauge choices). As a by product of the
Dirac string one can get the quantization condition for the electric charge. However,
as will be shown in section 5, the Dirac string is not necessary to obtain that
condition.

\section{EM duality without Dirac strings}

In this section we will present an alternative formulation of EM duality without the
use of Dirac strings. Besides $\phi_e$ and $\mathbf{A_m}$ defined in the previous
section, let us now introduce also a scalar potential $\phi_m$ associated to the
magnetic field and a vector potential $ \mathbf{A_e}$ associated to the electric
field. The electric field $\mathbf{E}$ and the magnetic induction $\mathbf{B}$ are
then expressed as:
\begin{equation}
\mathbf{E}=-\nabla\phi_e -\frac{\partial\mathbf{A_m}}{\partial
t}+\nabla\times\mathbf{A_e},
\end{equation}
\begin{equation}
\mathbf{B}=\nabla\phi_m +\frac{\partial\mathbf{A_e}}{\partial
t}+\nabla\times\mathbf{A_m}.
\end{equation}
Now we substitute these definitions in Maxwell's equation (2.14) and (2.15), and use
the Lorentz gauge given by:
\begin{equation}
\nabla\cdot\mathbf{A_m}+\frac{\partial\phi_e}{\partial t}=0,~~
\nabla\cdot\mathbf{A_e}+\frac{\partial\phi_m}{\partial t}=0.
\end{equation}
Thus we get
\begin{equation}
\begin{array}{l}
\frac{\partial^2}{\partial t^2} \phi_e -\nabla^2\phi_e=\rho_e ,\\
~\\
 \frac{\partial^2}{\partial
t^2}\mathbf{A_m}-\nabla^2\mathbf{A_m}=\mathbf{j_e}\\ ~~~~\\
\frac{\partial^2}{\partial t^2}\phi_m -\nabla^2\phi_m=-\rho_m,\\
 ~~~~\\
\frac{\partial^2}{\partial
t^2}\mathbf{A_e}-\nabla^2\mathbf{A_e}=-\mathbf{j_m}\\
\end{array}
\end{equation}

\noindent At this point we define
\begin{equation}
A_\mu^1=(\phi_e,-\mathbf{A_m}),~~~~ A^{\mu 1}
=(\phi_e,\mathbf{A_m}),
\end{equation}
\begin{equation}
A_\mu^2=(\phi_m,-\mathbf{A_e}),~~~~ A^{\mu
2}=(\phi_m,\mathbf{A_e}),
\end{equation}
and
\begin{equation}
J_\mu^1=J_\mu^e=(\rho_e, -\mathbf{j_e}),~~J_\mu^2=J_\mu^m=(\rho_m, -\mathbf{j_m}).
\end{equation}
Equations in (3.4) can then be written as
\begin{equation}
\partial_\mu\partial^\mu A_\nu^1=J_\nu^1
\end{equation}
\begin{equation}
\partial_\mu\partial^\mu A_\nu^2=-J_\nu^2
\end{equation}
In analogy to formula (2.6) we also write
\begin{equation}
F_{\mu\nu}^I=\partial_\mu A_\nu^I -\partial_\nu A_\mu^I,~~~~I=1,2.
\end{equation}
Taking into account that Lorentz gauge (3.3) can now be expressed as $$ \partial^\mu
A_\mu^I=0,$$ we have
\begin{equation}
\partial^\mu F_{\mu\nu}^I=g^{II'}J_\nu^{I'} ,
\end{equation}
where $$g^{II'}=\left(\begin{array}{cc}1&0\\0&-1\end{array}\right).$$

This is our first non trivial result. Equation (3.11) is the new version of Maxwell's
equations, which have manifest Lorentz symmetry and SO(2) dual symmetry. The currents
$J_\mu^I$ are conserved currents. In our approach this conservation reads
\begin{equation}
\partial^\nu J_\nu^I \propto\partial^\nu\partial^\mu
F_{\mu\nu}^I=0
\end{equation}
which are equivalent to equations (2.16) and (2.17).

We should mention that the general gauge transformation in this formalism is given by
$A^{I}_{\mu} \rightarrow A^{I}_{\mu}+ \partial_{\mu}\chi^{I}$. It is easy to check
that the fields $\mathbf{E}$, $\mathbf{B}$ as well as the field tensors in (3.10) and
Maxwell's equations (3.11) are all invariant under such a transformation.

Let us also stress that in the expressions above neither $F_{\mu\nu}^1$ nor
$F_{\mu\nu}^2$ have the same matrix form given by (2.11) and (2.12). In order to
facilitate comparison between our formalism and the one depicted in section 2 it is
useful to define
\begin{equation}
\mathcal{F}_{\mu\nu}=F_{\mu\nu}^1 +{^\ast} F_{\mu\nu}^2
\end{equation}
\begin{equation}
\widetilde{\mathcal{F}}_{\mu\nu}={^\ast} F_{\mu\nu}^1 -
F_{\mu\nu}^2
\end{equation}
where $\widetilde{\mathcal{F}}_{\mu\nu}$ can be viewed as a new
dual field tensor that, as we shall see, is specially adequate to
express the duality symmetry in a compact fashion. The matrix
form for $\mathcal{F}_{\mu\nu}$ is exactly the same as the form in
(2.11). On the other hand the newly introduced
$\widetilde{\mathcal{F}}_{\mu\nu}$ has a matrix form equal to the
one corresponding to ${^\ast} F_{\mu\nu}$ in the previous section
(equation (2.12)). Of course the electric and magnetic fields in
this new expressions are given by (3.1) and (3.2) whereas the
older ones were determined by (2.3) and (2.4). Since the vector
potentials in our formalism have no singularities one has
$\partial^\mu \,{^\ast}F_{\mu\nu}^I=0$, so Maxwell's equations
can also be written as
\begin{equation}
\begin{array}{l}
\partial^\mu \mathcal{F}_{\mu\nu}=\partial_\mu F_{\mu\nu}^1=J_{\nu}^1\\
\partial^\mu \widetilde{\mathcal{F}}_{\mu\nu}=-\partial_\mu F_{\mu\nu}^2=J_{\nu}^2
\end{array}
\end{equation}
This couple of equations is obviously invariant under the dual
transformation $
\mathcal{F}_{\mu\nu}\rightarrow\widetilde{\mathcal{F}}_{\mu\nu},~~
\widetilde{\mathcal{F}}_{\mu\nu}\rightarrow
-\mathcal{F}_{\mu\nu},~~J_{\mu}^1\rightarrow
J_{\mu}^2,~~J_{\mu}^2\rightarrow -J_{\mu}^1$, which are
equivalent to $\mathbf{E}\rightarrow-\mathbf{B}$ and
$\mathbf{B}\rightarrow\mathbf{E}$. This duality symmetry is
called special duality symmetry \cite{Donev}. In fact, (3.15) has
a more general duality symmetry under the following
transformations:

\begin{equation}
\left(\begin{array}{c}{\mathcal{F}'}_{\mu\nu}\\
\widetilde{\mathcal{F}}_{\mu\nu}'
\end{array}\right)=\left(\begin{array}{cc}a
&c\\b&d\end{array}\right) \left(\begin{array}{c}\mathcal{F}_{\mu\nu}\\
\widetilde{\mathcal{F}}_{\mu\nu}\end{array}\right),
\end{equation}

\begin{equation}
\left(\begin{array}{c}{J'}_{\mu}^{1}\\{J'}_{\mu}^{2}
\end{array}\right)=\left(\begin{array}{cc}a&c\\b&d
\end{array}\right)\left(\begin{array}{c}J_{\mu}^{1}
\\J_{\mu}^{2}\end{array}\right)
\end{equation}

\noindent with $ad-bc\neq0$. This symmetry should also hold for Maxwell's equations
(2.14) and (2.15). Indeed, invariance of these equations under the change

\begin{equation}
\left(\begin{array}{c}{\mathbf{E}'}\\ \mathbf{B}'
\end{array}\right)=\left(\begin{array}{cc}a&c\\b&d
\end{array}\right)\left(\begin{array}{c}\mathbf{E}
\\\mathbf{B}\end{array}\right)
\end{equation}

\noindent yields $a = d$ and $b = -c$. Moreover, if we impose that the energy density
and the Poynting vector are also invariant under this transformation we get $a^2 + b^2
=1$. It is then natural to introduce an angle $\alpha$ such that $a=\cos\alpha$ and
$b=\sin\alpha$. Hence the general duality transformation matrix coincides with the
general rotation matrix in two dimensions. Thus it becomes apparent that the general
EM duality symmetry is the SO(2) symmetry.

Let us also mention that it would be desirable to get equation (3.15) from a
variational principle, exactly as it is done in the absence of monopoles \cite{Weyl}
(see also \cite{Finzi}). This issue will be considered in section 6.

It is also important to stress that our two potentials formulation can be recast in
terms of one unique vector field $\mathcal{A}_\mu$ defined as
\begin{equation}
\mathcal{A}_\mu=A_\mu^1+{^\ast A}_\mu^2,
\end{equation}
where ${^\ast A}_\mu^I$ are defined through
\begin{equation}
\partial_\mu
{^\ast}A_\nu^I=\frac{1}{2}\epsilon_{\mu\nu}^{~~\alpha\beta}\partial_\alpha A_\beta^I .
\end{equation}
The field tensor $\mathcal{F}_{\mu\nu}$ can then be expressed as
\begin{equation}
\mathcal{F}_{\mu\nu}
=\partial_\mu\mathcal{A}_\nu-\partial_\nu\mathcal{A}_\mu .
\end{equation}
This tensor has the same form as in equation (2.6), but here the dual of
$\mathcal{F}_{\mu\nu}$ coincides with $\widetilde{\mathcal{F}}_{\mu\nu}$, it does not
satisfy an equation of the form (2.9). As a consequence the equations in (2.10) are
now replaced by equations (3.15). Let us also stress that after quantization the field
$\mathcal{A}_\mu$ will be associated to the photon.

\section{ Application to specific electromagnetic systems}

In this section we will give some explicit solutions for specific
static systems in the presence of both electric and magnetic
sources. At the end we will briefly comment on the formal
solutions for the general non-static case.

In a static situation the Maxwell's equation (3.4) becomes
\begin{equation}
\begin{array}{l}
\nabla^2\phi_e=-\rho_e (\mathbf{x}),~~~~\nabla^2\phi_m=\rho_m
(\mathbf{x}),\\ \nabla^2
\mathbf{A_m}=-\mathbf{J_e}(\mathbf{x}),~~~~
\nabla^2\mathbf{A_e}=\mathbf{J_m}(\mathbf{x})
\end{array}
\end{equation}
For simplicity, we shall consider a dyon with electric charge $q$ and magnetic charge
$g$ placed at the origin of the coordinate system:
\begin{equation}
\begin{array}{l}
\nabla^2\phi_e=-q\,\delta (\mathbf{x}),~~~~\nabla^2\phi_m=g\,\delta (\mathbf{x}),\\
\nabla^2 \mathbf{A}_I=0,~~~~~~ I = e, m.
\end{array}
\end{equation}
The solutions in boundless space are
\begin{equation}
\begin{array}{l}
\phi_e=\displaystyle\frac{1}{4\pi }\frac{q}{r},~~~~\phi_m=-\displaystyle\frac{1}{4\pi
}\frac{g}{r},\\~~~\\ \mathbf{A}_I= 0,~~~~~~ I = e, m.
\end{array}
\end{equation}
Then from (3.1) and (3.2), the field strengths are given by
\begin{equation}
\begin{array}{l}
\mathbf{E}=\displaystyle\frac{q\,\mathbf{r}}{4\pi r^3}\\ ~~~\\
\mathbf{B}=\displaystyle\frac{g\,\mathbf{r}}{4\pi r^3}.
\end{array}
\end{equation}
Let us emphasize that these simple solutions are obtained without using the concept of
Dirac string.

As a second example we now consider a steady current of dyons in a circular loop. To
be definite let us place the circle with radius $R$ in the $xy$ plane and use
spherical coordinates $r$, $\theta$ and $\varphi$. The only non-vanishing components
of the current densities correspond to the $\varphi$ direction:
\begin{equation}
J_\varphi^I(r,\theta)=J^I\delta (\cos (\theta ))\frac{\delta (r-R)}{R},~~~~~~ I = e,
m.
\end{equation}

\noindent where $J$ is the number of dyons passing through a cross-section in the unit
of time, $J^e =qJ$ and $ J^m =gJ$. Since $J^I$ has only $\varphi$-components the same
will happen with $\mathbf{A}^I$. Therefore the solutions of Maxwell's equation (4.1)
for this case read
\begin{equation}
 A_\varphi^I (r,\theta)=\frac{J^I R}{\pi \sqrt{R^2 +r^2 +2Rr\sin\theta}}\left(\frac{(2-k^2 )K(k)-2E(k)}
 {k^2}\right)
\end{equation}
where $$k^2 =\frac{4Rr\sin\theta}{R^2 +r^2 +2Rr\sin\theta}$$ and $ E(k)$ and $K(k)$
are elliptic integrals.

For small $k$, one gets
\begin{equation}
A_\varphi^I=\frac{J^I R^2}{4}\frac{r\sin\theta}{(R^2 +r^2 +2Rr\sin\theta)^{3/2}}
\end{equation}
\noindent Noticing that in this example one has $\phi_e=\phi_m=0$, we obtain
$\mathbf{E}=\nabla\times\mathbf{A_e},\mathbf{B}=\nabla\times\mathbf{A_m} $. This means
that for $k$ small (which corresponds to $R>>r, or~ r>> R, or~ \theta <<1 $), we find
\begin{equation}
\begin{array}{l}
B_r=\displaystyle\frac{qJ\cos\theta}{4}\frac{2R^2+2r^2+Rr\sin\theta}{(R^2+r^2+2Rr\sin\theta)^{5/2}}\\
~~~\\
B_\theta=-\displaystyle\frac{qJ\sin\theta}{4}\frac{2R^2+2r^2+Rr\sin\theta}{(R^2+r^2+2Rr\sin\theta)^{5/2}}\\
~\\ B_\varphi=0
\end{array}
\end{equation}
and
\begin{equation}
\begin{array}{l}
E_r=\displaystyle\frac{gJ\cos\theta}{4}\frac{2R^2+2r^2+Rr\sin\theta}{(R^2+r^2+2Rr\sin\theta)^{5/2}}\\
~~~\\
E_\theta=-\displaystyle\frac{gJ\sin\theta}{4}\frac{2R^2+2r^2+Rr\sin\theta}{(R^2+r^2+2Rr\sin\theta)^{5/2}}\\
~\\ E_\varphi=0 .
\end{array}
\end{equation}
When the magnetic charge vanishes ($g=0$), the above results coincide with the
well-known expressions obtained for an electric charge moving steadily in a circular
loop. On the other hand, for $g\neq 0$, it is not simple to evaluate the fields using
the concept of Dirac strings.

For completeness let us now display the formal solutions for
general non-static distributions:
\begin{equation}
\rho_I=\rho_I(\mathbf{x,t}),~~ \mathbf{J}_I=\mathbf{J}_I
(\mathbf{x,t}),~~~~ I = 1,2. ~{\rm represent }~ I= e, m.
\end{equation}
In principle this problem can be analyzed by means of the retarded potential method,
exactly as it is done in $g=0$ classical electrodynamics \cite{Jackson}. The solution
of equations (3.4) is then given by
\begin{equation}
\begin{array}{l}
\phi_I (\mathbf{x},t)
=\displaystyle\frac{1}{4\pi}\int\frac{g^{II'}\rho_{I'}
(\mathbf{x}',t-r)}{r}\,d^3 x'\\ ~\\ \mathbf{A}_I (\mathbf{x},t)
=\displaystyle\frac{1}{4\pi}\int\frac{g^{II'}\mathbf{J}_{I'}
(\mathbf{x}',t-r)}{r}\,d^3 x',
\end{array}
\end{equation}
\noindent where $r=|\mathbf{x}-\mathbf{x}'|$ and $g^{II'}$ is
given by the definition below the equation (3.11), then the field
strengths are

\begin{equation}
\begin{array}{l}
\mathbf{E}(\mathbf{x},t) =\displaystyle\frac{1}{4\pi}\int\rho_e
(\mathbf{x}',t-r)\frac{\mathbf{r}}{r^3}\,d^3x'\\~\\
+\displaystyle\frac{1}{4\pi}\int\mathbf{J}_m
(\mathbf{x}',t-r)\,{\bf
\times}\,\frac{\mathbf{r}}{r^3}\,d^3x'-\displaystyle\frac{1}{4\pi}\int\frac{1}{r}\frac{\partial\mathbf{J}_m
(\mathbf{x}',t-r)}{\partial t}\,d^3x' ,
\end{array}
\end{equation}
\noindent and
\begin{equation}
\begin{array}{l}
\mathbf{B}(\mathbf{x},t) =\displaystyle\frac{1}{4\pi}\int\rho_m
(\mathbf{x}',t-r)\frac{\mathbf{r}}{r^3}\,d^3x'\\~\\ -
\displaystyle\frac{1}{4\pi}\int\mathbf{J}_e
(\mathbf{x}',t-r)\,{\bf \times}\,\frac{\mathbf{r}}{r^3}\,d^3x'
+\displaystyle\frac{1}{4\pi}\int\frac{1}{r}\frac{\partial\mathbf{J}_e
(\mathbf{x}',t-r)}{\partial t}\,d^3x' .
\end{array}
\end{equation}

The solutions for the general static case is obtained from the
expressions above  by omitting the time derivative terms and of
course one shoud also omit $t$ and $t-r$ in the corresponding
arguments.

\section{Generalized Lorentz force and the electric charge quantization condition}

As it is well-known a particle with electric charge $q$ moving in the electromagnetic
field will be subjected to a Lorentz force:
\begin{equation}
\mathbf{F}=q\,\mathbf{E}+q\,\mathbf{v}\times\mathbf{B},
\end{equation}
\noindent where $\mathbf{v}$ is the velocity of the particle. From the dual symmetry,
we know that if a particle with magnetic charge $g$ moves in an electromagnetic field,
it will also feel a force given by
\begin{equation}
\mathbf{F}=g\,\mathbf{B}-g\,\mathbf{v}\times\mathbf{E}
\end{equation}
Now, if we consider a dyon moving in the electromagnetic field, it will gain a
generalized Lorentz force:
\begin{equation}
\mathbf{F}=q\,(\mathbf{E}+\mathbf{v}\times\mathbf{B})+g\,(\mathbf{B}-\mathbf{v}\times\mathbf{E})
\end{equation}
\noindent and the equation of motion in a covariant form reads

\begin{equation}
m\,\ddot{x^\mu}=(q\,\mathcal{F}^{\mu\nu}
+g\,\widetilde{\mathcal{F}}^{\mu\nu})\,\dot{x_\nu}.
\end{equation}
When $g=0$, we have $\mathcal{F}_{\mu\nu}\rightarrow F_{\mu\nu}$,
equation (5.4) returns to the well know result in electrodynamics.


Now let us consider a particle with electric charge $q$ moving in the field produced
by a monopole with magnetic charge $g$. We place a very heavy  monopole at the origin
of the coordinate system and use for $\mathbf{B}$ the solution obtained in the
previous section (for the static case), such that the equation of motion of the
electric charge is

\begin{equation}
m\,\frac{d^2 \mathbf{x}}{d
t^2}=\frac{q\,g}{4\pi}\,\mathbf{v}\times\frac{\mathbf{r}}{r^3}.
\end{equation}

It is easily shown that the kinetic energy and the total angular momentum are
constants of motion of this system. Indeed, through elementary manipulations one can
show that $$\frac{d}{d t} (\mathbf{L}-\kappa \hat{\mathbf{r}})=0 , $$ where $\kappa =
\frac{q g}{4 \pi}$ and $\mathbf{L}=\mathbf{r}\times m\mathbf{v}$ is the orbital
angular momentum. Thus we find the following constant of motion,
\begin{equation}
\mathbf{J}=\mathbf{L}-\kappa\hat{\mathbf{r}}
\end{equation}

This means that the total angular momentum of the system must be identified with
$\mathbf{J}$ whereas the second term $(-\kappa\hat{\mathbf{r}})$ is nothing but the
angular momentum of the electromagnetic field. Since we have decomposed $\mathbf{J}$
into two orthogonal parts one immediately obtains
\begin{equation}
\mathbf{J}^2=\mathbf{L}^2 +\kappa^2
\end{equation}

\noindent and now it is obvious that quantization of the total and orbital angular
momentum (via equation (5.6)) translates into the Dirac quantization condition:
\begin{equation}
\kappa =\frac{1}{2}\, \hbar ~~\Rightarrow~~ q\,g=2\pi n,~~~~~~~~~ n=1,2,\cdots~~(\hbar
=1).
\end{equation}
An immediate corollary is that if a single monopole with magnetic charge $g$ exists,
then the electric charge is quantized in units of $2\pi/g$. If we consider a dyon with
electric and magnetic charge $(q_1 ,g_1)$ moving in the field of another dyon with
charges $(q_2, g_2)$, we can derive the new electromagnetic angular momentum which
reads
\begin{equation}
\mathbf{L_{em}}=\frac{q_2g_1-q_1g_2}{4\pi}~\,\frac{\mathbf{r}}{r}
\end{equation}
and the corresponding quantization condition is
\begin{equation}
q_1g_2-q_2g_1=2\pi n,~~~~~~~~~~ n=1,2,\cdots
\end{equation}

Therefore we were able to obtain the electric charge quantization condition without
using the concept of Dirac string.

\section{Final remarks}
We end this work by making some remarks concerning the Lagrangian formulation.
According to the description of EM duality presented in this paper, the field
equations are given by equation (3.11). These equations can be derived from the
following Lagrangian density
\begin{equation}
\begin{array}{lll}
\mathcal{L}&=&\frac{1}{4}(\widetilde{\mathcal{F}}_{\mu\nu})^2-\mathcal{A}_\mu J^{\mu
1}- \widetilde{\mathcal{A}}_\mu J^{\mu 2},
\end{array}
\end{equation}
where $\mathcal{A}_\mu$ is given by (3.19),
$\widetilde{\mathcal{A}}_\mu=A_\mu^2-{^\ast}A_\mu^1$,
$\widetilde{\mathcal{F}}_{\mu\nu}=-(\partial_\mu \widetilde{\mathcal{A}}_\nu
-\partial_\nu \widetilde{\mathcal{A}}_\mu )$ and we have used $
(\mathcal{F}_{\mu\nu})^2=-(^\ast\mathcal{F}_{\mu\nu})^2=-(\widetilde{\mathcal{F}}_{\mu\nu})^2$.
Employing the identities
 $$
\frac{\partial \mathcal{F}^2}{\partial (\partial_\mu
\mathcal{A}_\nu)}=4\mathcal{F}^{\mu\nu}$$

$$ \frac{\partial \widetilde{\mathcal{F}}^2}{\partial (\partial_\mu
\widetilde{\mathcal{A}}_\nu)}=-4\widetilde{\mathcal{F}}^{\mu\nu}, $$ it is easy to
show that the corresponding Euler-Lagrange equations $$\partial_\mu (
\frac{\partial\mathcal{L}}{\partial
(\partial_\mu\mathcal{A}_\nu)})=\frac{\partial\mathcal{L}}{\partial\mathcal{A}_\nu}$$
 $$\partial_\mu (\frac{ \partial\mathcal{L}}{\partial
(\partial_\mu\widetilde{\mathcal{A}}_\nu)})=\frac{\partial\mathcal{L}}
{\partial\widetilde{\mathcal{A}}_\nu }$$ give the right Maxwell's equation (3.15).
Following the above depicted calculation one can see that the two-potential formalism
proposed in this work is equivalent to the usual one potential one, which is
consistent with the existence of only one kind of photon in the real world.

The first term of equation (6.1) is associated to the free electromagnetic field,
whereas the second and third terms correspond to the interaction between field and
source (currents) and the interaction between electric source and magnetic source.
More explicitly, the interaction between electric and the magnetic sources can be
expressed as
\begin{equation}
-{^\ast}A_\mu^2J^{\mu 1}+{^\ast}A_\mu^1J^{\mu 2}.
\end{equation}
 Taking into account equation (3.20), we can formally write
 $$
{^\ast}A_\mu^I=-\frac{1}{2}\epsilon_{\mu\nu\alpha\beta}\int_P^x
\partial^\alpha A^{\beta I} dx^\nu
 $$
 and using (4.11) (for the static case) we have
$$
 A_\mu^I (x)=\frac{1}{4\pi}\int \frac{g^{II'}J_\mu^{I'} (\mathbf{x'})}{r}d^3x'.
$$

Therefore we have, \bea {^\ast}A_\mu^1J^{\mu
2}=-\frac{1}{2}\epsilon_{\mu\nu\alpha\beta}\int_P^x
\partial^\alpha A^{\beta 1}(y) dy^\nu\cdot J^{\mu 2}(x) \nonumber \\
=-\frac{1}{8\pi}\epsilon_{\mu\nu\alpha\beta}\int_\Omega J^{\beta
1}(\mathbf{x'})[\int_P^x \frac{\partial}{\partial
y_\alpha}\frac{1}{|\mathbf{y}-\mathbf{x'}|}dy^\nu]J^{\mu
2}(\mathbf{x})d^3x'\nonumber\\ \eea which means that ${^\ast}A_\mu^1J^{\mu 2}$ is just
the non local interaction between the electric current $ J^{\beta 1}$ and the magnetic
current $ J^{\mu 2}$. One has a similar result for the term ${^\ast}A_\mu^2 J^{\mu 1}$
which is obtained by interchanging $J^{\mu 1}$ and $J^{\mu 2}$ in the above equation.

\vspace{0.5cm}

In summary we have presented an alternative description of classical electromagnetism.
Relevant features of this formulation are its manifest Lorentz covariance and its
simple realization of duality symmetry. However, the main advantage is the fact that
the four-vector potentials introduced in our formalism have no singularities, thus
allowing a description of dyonic dynamics without Dirac strings. We think that these
properties will be helpful when carrying over the quantization of the theory. We hope
to report on this issue in a forthcoming article \cite{LN}.

\paragraph{Acknowledgements}
We thank S. Antoci for sending us a copy of Ref. \cite{Finzi}. We are also grateful to
S. Carneiro for useful criticisms. This work was supported by the Consejo Nacional de
Investigaciones Cient\'{\i}ficas y T\'ecnicas (CONICET), Argentina. The authors also
recognize the support of the Third World Academy of Sciences (TWAS).

\end{document}